\documentclass[galaxies,article,accept,pdftex,moreauthors]{Definitions/mdpi} 
\firstpage{1} 
\pubvolume{1}
\issuenum{1}
\articlenumber{0}
\pubyear{2023}
\copyrightyear{2023}
\usepackage{lscape}
\externaleditor{{Academic Editor: Roberta M. Humphreys} 
}
\datereceived{15 March 2023} 
\daterevised{6 June 2023} % Only for the journal Acoustics
\dateaccepted{8 June 2023} 
\datepublished{date} 
\usepackage{lscape}
\externaleditor{{Academic Editor: Roberta M. Humphreys} 
}
\datereceived{15 March 2023} 
\daterevised{6 June 2023} % Only for the journal Acoustics
\dateaccepted{8 June 2023} 
\datepublished{date} 
\hreflink{https://doi.org/} % If needed use \linebreak
\newcommand{\msol}{\mbox{M$_{\odot}$}}

\Title{Discovering new B[e] supergiants and candidate Luminous Blue Variables in nearby galaxies}

\TitleCitation{Discovering new B[e] supergiants and candidate Luminous Blue Variables in nearby galaxies}

\Author{Grigoris Maravelias$^{1,2\dagger}$\orcidA{}, Stephan de Wit$^{1,3}$\orcidB{},
Alceste Z. Bonanos$^{1}$\orcidC,
Frank Tramper$^{4}$\orcidD,
Gonzalo Munoz-Sanchez$^{1,3}$\orcidE,
Evangelia Christodoulou$^{1,3}$\orcidF}

\AuthorNames{Grigoris Maravelias, Stephan de Wit, Alceste Z. Bonanos, Frank Tramper, Gonzalo Munoz-Sanchez and Evangelia Christodoulou}

\AuthorCitation{Maravelias, G.; de Wit, S.;
Bonanos, A. Z.; Tramper, F.; Munoz-Sanchez, G.; Christodoulou, E.}
\address{%
$^{1}$ \quad IAASARS, National Observatory of Athens, GR-15326 Penteli, Greece\\
$^{2}$ \quad Institute of Astrophysics FORTH, GR-71110, Heraklion, Greece\\
$^{3}$ \quad Department of Physics, National and Kapodistrian University of Athens,  Panepistimiopolis, Zografos, GR-15784, Greece\\
$^{4}$ \quad Institute of Astronomy, KU Leuven, Celestijnlaan 200D, 3001 Leuven, Belgium}

\corres{Correspondence: maravelias@noa.gr}

\abstract{Mass loss is one of the key parameters that determine stellar evolution. Despite the progress we have achieved over the last decades we still cannot match the observational derived values with theoretical predictions. Even worse, there are certain phases, such as the B[e] supergiants (B[e]SGs) and the Luminous Blue Variables (LBVs), where significant mass is lost through episodic or outburst activity. This leads to various structures around them that permit dust formation, making these objects bright IR sources. The ASSESS project aims to determine the role of episodic mass in the evolution of massive stars, by examining large numbers of cool and hot objects (such as B[e]SGs/LBVs). For this, we initiated a large observing campaign to obtain spectroscopic data for $\sim1000$ IR selected sources in 27 nearby galaxies. Within this project we successfully identified 7 B[e] supergiants (one candidate) and 4 Luminous Blue Variables of which 6 and 2, respectively, are new discoveries. We used spectroscopic, photometric, and light curve information to better constrain the nature of the reported objects. We particularly note the presence of B[e]SGs at metallicity environments as low as 0.14 Z$_{\odot}$. }

\begin{document}

\section{Introduction}
How \textit{exactly} single massive stars, born as O/B-type main-sequence stars, progress to more evolved phases and eventually die remains an open question. Binarity, which has an important implication in the evolution, even further complicates the quest for an answer. Observational data has revealed a number of transitional phases in which massive stars can be found, also known as the massive star "zoo". Whether they pass through certain phases or not depends on their initial mass ($\geq$ 8~\msol), metallicity ($Z$), rotational velocity ($v_\textrm{rot}$), mass loss properties and binarity \cite{Ekstrom2012, Georgy2013, Smith2014, Eldridge2022}. Although some of them are quite distinct (e.g. Wolf-Rayet stars as opposed to Red Supergiants - RSGs), there are phases which display common observables, such as B[e] supergiants (B[e]SGs) and Luminous Blue Variables (LBVs).

The B[e] phenomenon is characterized by numerous emission lines in the optical spectra \cite{Lamers1998}. In particular, there is strong Balmer emission, low excitation permitted (e.g., Fe~\textsc{ii}), and forbidden lines (of [Fe~\textsc{ii}], and [O~\textsc{i}]), as well as strong near- or mid-IR excess due to hot circumstellar dust. However, this can be observed in sources at different evolutionary stages (such as in Herbig AeBe stars, symbiotic systems, and compact planetary nebulae, see~\cite{Lamers1998} for detailed classification criteria). The B[e]SGs form a distinct subgroup based on a number of secondary criteria. They are luminous stars (log~$L/L_{\odot}\gtrsim 4.0$), showing broad Balmer emission lines with P Cygni or double-peaked profiles. They may also display evidence of chemically processed material (e.g., $^{13}$CO enrichment, TiO) which points to an evolved nature, although it is not yet certain if they are in pre- or post-RSG phases \cite{Kraus2009, Kraus2019a}. The presence of the hot circumstellar dust is due to a complex circumstellar environment (CSE) formed by two components, a stellar wind radiating from the poles and a denser equatorial ring-like structure \citep{Zickgraf1985, Zickgraf1989, Aret2012, deWit2014, Maravelias2018}. However, the formation mechanism of this structure remains elusive. A variety of mechanisms have been proposed, such as the following: fast rotation \cite{Kraus2006}, the bi-stability mechanism \cite{Petrov2016}, slow-wind solutions \cite{Cure2005}, magneto-rotational instability \cite{Krticka2015} , mass transfer in binaries \cite{Wheelwright2012b}, mergers \cite{Podsiadlowski2006}, non-radial pulsations or the presence of objects that clear their paths \cite{Kraus2016}. Although poorly constrained, their initial masses range from roughly 10~\msol\, to less than 40~\msol\, (Mehner 2023, IAU S361, subm.). 

The LBVs are another rare subgroup of massive evolved stars, considered to represent a transitional phase from  massive O-type main-sequence to Wolf--Rayet stars (e.g., \cite{Humphreys2014, Smith2014, Humphreys2017, Weis2020}). They experience instabilities that lead to photometric variability, {{typically referred to as S Dor cycles} \cite{Weis2020}}, as well as outbursts and episodic mass loss, {similar to the giant eruption of $\eta$ Carina that resulted in large amounts of mass lost through ejecta (e.g., \cite{Vink2012}}). {It is not yet fully understood whether these two types of variability are related (e.g., \cite{Davidson2020}}). Apart from the evident photometric variability, their spectral appearance changes significantly during their outburst activities ({S Dor cycle}). It is typical to experience loops from hot (spectra of O/B type) to cool states (A/F spectral types while in outbursts). Depending on the luminosity, the brightest LBVs (log~$L/L_\odot>5.8$) seem to directly originate from main-sequence stars (with mass $>50$~\msol), while the less luminous ones are possibly post-RSG objects that have lost almost half of their initial masses (within the range of $\sim$25--40~\msol) during the RSG phase (Mehner 2023, IAU S361, subm.). Currently, various mechanisms have been suggested, such as radiation and pressure instabilities, stellar rotation, and binarity (see the reviews on the theory and observational evidence in \cite{Davidson2020, Weis2020}, Mehner 2023, IAU S361, subm., and the references therein) and, as such, no comprehensive theory exists to explain them.   

Therefore, if and how these two phases are linked remains an open question. B[e]SGs tend to have initial masses with a wide range below the most luminous LBVs, and in accordance with the less luminous ones. The presence of similar lines in their spectra points to similarities in their CSEs, with shells and bipolar nebulae observed in both cases~\cite{Wachter2010, Weis2020, Liimets2022}.

Due to their photometric variability, LBVs are more commonly detected in other galaxies compared to B[e]SGs, which generally display less variability.\endnote{ \citet{Lamers1998} commented on a relative low variation of up to 0.2 mag, which was not the case in more recent studies, see Section~\ref{s:light_variability} for more details. }.  
Therefore, B[e]SGs need to be searched for to be discovered. This has only been successful for 56 (candidate) sources in the Galaxy and for the Magellanic Clouds (MCs), M31 and M33, and M81 \citep{Kraus2019a}, and only recently in NGC 247 \cite{Solovyeva2020}. On the other hand, LBVs have been found in more galaxies (additional to the aforementioned), such as IC 10, IC 1613, NGC 2366, NGC 6822, NGC 1156, DDO 68, and PHL 293B \citep{Richardson2018, Wofford2020, Guseva2022, Solovyeva2023, Weis2020}, summing up to about 150 sources (including candidates).    

This paper presents the discovery of new B[e]SGs and LBV candidates found with a systematic survey to identify massive, evolved, dusty sources in nearby galaxies ($\leq$5~Mpc), as part of the ASSESS project\endnote{\highlighting{\url{https://assess.astro.noa.gr}}} (Bonanos 2023, IAU S361, subm.). In Section \ref{sec2} we provide a short summary of the observations and of our approach, in Section \ref{sec3} we present the new sources, and in Sections \ref{s:discussion} and \ref{sec5} we discuss and conclude our work.

%%%%%%%%%%%%%%%%%%%%%%%%%%%%%%%%%%%%%%%%%%
\section{Materials and Methods}\label{sec2}

\subsection{Galaxy Sample}

For the ASSESS project, a list of 27 nearby galaxies ($\le$5~Mpc) was compiled (see Bonanos 2023, IAU S361, subm.). In this paper, we present our results from a sub-sample of these galaxies (Table~\ref{t:galaxies}) for which the spectral classification is final, while for another set we have scheduled observations in queue and have submitted proposals. For some galaxies (e.g., MCs) data have been collected through other catalogs/surveys and are presented separately (e.g., \citep{deWit2022, Yang2019, Yang2020, Yang2021}). 

\begin{table}[H]
%\footnotesize%\scriptsize\footnotesize
\caption{Properties of galaxies examined in this work: galaxy ID (column 1), sky coordinates (columns 2 and 3), galaxy type (column 4), distance (column 5), metallicity (column 6), and radial velocity (RV, column 7). % GM better description of the table
\label{t:galaxies}}
%\newcolumntype{C}{>{\centering\arraybackslash}X}
%\begin{adjustwidth}{-\extralength}{0cm}
%\newcolumntype{C}[1]{>{\PreserveBackslash\centering}m{#1}}
%\newcolumntype{R}[1]{>{\PreserveBackslash\raggedleft}m{#1}}
%\newcolumntype{L}[1]{>{\PreserveBackslash\raggedright}m{#1}}

\setlength{\tabcolsep}{1.8mm}
\resizebox{\textwidth}{!}{
\begin{tabularx}{\textwidth}{lllllrr}
\toprule
\textbf{ID}	& \textbf{R.A.}	& \textbf{Dec.} &  \textbf{Gal. Type} & 
\textbf{Distance} & \textbf{Metal.~}\boldmath{$^1$} & \textbf{RV~}\boldmath{$^{1,2}$} \\
            & \textbf{(J2000)} &\textbf{ (J2000)} & & \textbf{(Mpc)} & \boldmath{$(Z_\odot)$} & \textbf{(km~s}\boldmath{$^{-1}$})\\
\textbf{(1)} & \textbf{(2)} & \textbf{(3)} & \textbf{(4)} &\textbf{ (5)} & \textbf{(6)} & \textbf{(7)} \\
\midrule

WLM	      & 00:01:58 & $-$15:27:39 & SB(s)m: sp & 0.98 $\pm$ 0.04 {\cite{Zgirski2021}} & 0.14 {\cite{Urbaneja2008}} & $-$130 $\pm$ 1 {\cite{McConnachie2012}} \\
NGC 55    & 00:14:53 & $-$39:11:48 & SB(s)m: sp & 1.87 $\pm$ 0.02 {\cite{Zgirski2021}} & 0.27 {\cite{Hartoog2012}} & 129 $\pm$ 2 {\cite{McConnachie2012}} \\
IC 10     &	00:20:17 & +59:18:14 & dIrr IV/BCD & 0.80 $\pm$ 0.03 {\cite{Sanna2008}} & 0.45 {\cite{Tehrani2017}} & $-$348 $\pm$ 1 {\cite{McConnachie2012}}  \\
NGC~247	  & 00:47:09 & $-$20:45:37 & SAB(s)d    & 3.03 $\pm$ 0.03 {\cite{Zgirski2021}} & 0.40 {\cite{Kacharov2018}} & 156 {\cite{Tully2016}} \\
NGC~253   & 00:47:33 & $-$25:17:18 & SAB(s)c    & 3.40 $\pm$ 0.06 {\cite{Madore2020}} & 0.72 {\cite{Spinoglio2022}} & 259 {\cite{Meyer2004}}\\
NGC~300   & 00:54:53 & $-$37:41:04 & SA(s)d     & 1.97 $\pm$ 0.06 {\cite{Zgirski2021}} & 0.41 {\cite{Kudritzki2008}} & 146 $\pm$ 2 {\cite{McConnachie2012}} \\ 
NGC 1313 & 03:18:16& $-$66:29:54 & SB(s)d & 4.61 $\pm$ 0.17 {\cite{Qing2015}} & 0.57 {\cite{Hernandez2022}} & 470 {\cite{Koribalski2004}} \\
NGC~3109  & 10:03:07 & $-$26:09:35 & SB(s)m edge-on & 1.27 $\pm$ 0.03 {\cite{Zgirski2021}} & 0.21 {\cite{Hosek2014}} & 403 $\pm$ 2 {\cite{McConnachie2012}}  \\
Sextans~A & 10:11:01 & $-$04:41:34 & IBm        & 1.34 $\pm$ 0.02 {\cite{Tammann2011}} & 0.06 {\cite{Kniazev2005}} & 324 $\pm$ 2 {\cite{McConnachie2012}}  \\
M83       & 13:37:01 & $-$29:51:56 & SAB(s)c    & 4.90 $\pm$ 0.20 {\cite{Bresolin2016}} & 1.58 {\cite{Hernandez2019}} & 519 {\cite{Meyer2004}} \\
NGC~6822  & 19:44:58 & +14:48:12 & IB(s)m     & 0.45 $\pm$ 0.01 {\cite{Zgirski2021}} & 0.32 {\cite{Dopita2019}} & $-$57 $\pm$ 2 {\cite{McConnachie2012}} \\
NGC~7793  & 23:57:50 & $-$32:35:28 & SA(s)d     & 3.47 $\pm$ 0.04 {\cite{Zgirski2021}} & 0.42 {\cite{DellaBruna2021}} & 227 {\cite{Meyer2004}}  \\
% \\

\bottomrule
\end{tabularx}}
\footnotesize{$^1$ The numbers presented here reflect the mean value per galaxy. $^2$ The RV errors correspond to the statistical error and not the systemic one, which is (typically)  larger.}
%\end{adjustwidth}
\end{table}

The aim of the ASSESS project is to determine the role of episodic mass loss by detecting and analyzing dusty evolved stars that are primary candidates to exhibit episodic mass loss events (Bonanos 2023, IAU S361, subm.). This mass loss results in the formation of complex structures, such as shells and bipolar nebulae in Wolf--Rayet stars and LBVs (e.g.,~\cite{Gvaramadze2010, Wachter2010}), detached shells in AGBs and RSGs (e.g.,~\cite{Cox2012}), disks and rings around B[e]SGs (e.g.,~\cite{deWit2014, Kraus2019a}, or even the dust-enshrouded shells within which the progenitors of Super-Luminous Supernovae lay (e.g., \cite{Smith2011, Zhang2012, GalYam2019}). The  presence of these dusty CSEs makes these sources bright in mid-IR imaging. Therefore, we based our catalog construction on published point-source \textit{Spitzer} catalogs \citep{Spitzer2004}. Since IR data alone cannot distinguish between these sources, the base catalogs were supplemented with other optical and near-IR surveys (Pan-STARRS1;~\cite{PanSTARRS2016}, VISTA Hemisphere Survey---VHS;~\cite{VISTA2013}, \textit{Gaia} DR2; \cite{Gaia2016mission,Gaia2018dr2}). \textit{Gaia} information was also used to remove foreground sources when possible (see \cite{Maravelias2022}, and Tramper et al., in prep., for more details).  

Given this data collection, we performed a selection process to minimize contamination by AGB stars and background IR galaxies/quasars. An absolute magnitude cut of M$_{[3.6]}\le-9.0$ \citep{Yang2020} and an apparent magnitude cut at m$_{[4.5]}\le15.5$ \citep{Williams2015} were applied to avoid AGB stars and background galaxies, respectively. In order to select the dusty targets we considered all sources with an IR excess, defined by the color term $\textrm{m}_{[3.6]}-\textrm{m}_{[4.5]}>0.1$~mag (to exclude the majority of foreground stars, for which this is approximately 0, and to select the most dusty IR sources). The three aforementioned criteria served as a minimum to consider a source as a priority target. Consequently, the reddest and brightest point-sources in the \textit{Spitzer} catalogs were given the highest priority. An extensive priority list/system was constructed by imposing certain limits for the color term, M$_{[3.6]}$, and the presence of an optical counterpart (for more details, see Tramper et al., in prep.). Depending on the galaxy size we ended up with a few tens to hundreds of targets per galaxy.

To obtain spectroscopic data for such a large number of targets we required instruments with multi-object spectroscopic modes. With these we could allocate up to a few tens of objects per pointing. Multiple pointings (with dithering and/or overlap) were applied to cover more extended galaxies and when the density of the target was high. Therefore, when we were creating the necessary multi-object masks we were forced to select sources based on the spatial limitations (e.g., located out of the field-of-view or at the sensor's gap) and spectral overlaps. Consequently, some priority targets were dropped and, additionally, non-priority targets (``fillers'', i.e., sources dropped through the target selection approach described previously) were added to fill the space.

\subsection{Observations}

To verify the nature of our selected targets we needed spectroscopic information. Since this is not available for the majority of the ASSESS galaxies, we initiated an observation campaign to obtain low resolution spectra. Given the large number of targets, along with the sizes of the galaxies, we used the multi-object spectroscopic modes of the Optical System for Imaging and low-Intermediate-Resolution Integrated Spectroscopy (OSIRIS;~\cite{OSIRIS}), on the 10.4 m GTC (\cite{GTC}, for the galaxies visible from the Northern hemisphere, i.e., IC 10 and NGC 6822). We used the FOcal Reducer/low dispersion Spectrograph 2 (FORS;~\cite{FORS2}), at 8.2 m ESO/VLT (for the Southern galaxies, i.e., the rest of Table 1). The resolving power and wavelength coverage was similar for both instruments, at $\sim$500--700 over the range \mbox{$R\sim$5300--9800~\AA\,} for GTC/OSIRIS and $R\sim$1000 over the range $\sim$5200--8700~\AA\, for VLT/FORS2. Details for the observations and data reduction can be found at Munoz-Sanchez et al., in prep., for the GTC/OSIRIS campaign and Tramper et al., in prep., for the VLT/FORS2 campaign. Here we provide only a short overview of the data reduction followed. 

For the OSIRIS data we used the \textit{GTCMOS} package\endnote{\url{https://www.inaoep.mx/~ydm/gtcmos/gtcmos.html} accessed 1/9/2022---śee also \cite{GomezGonzalez2016} } which is an \textsc{IRAF}\endnote{\textsc{IRAF} is distributed by the National Optical Astronomy Observatory, which is operated by the Association of Universities for Research in Astronomy (AURA) under cooperative agreement with the National Science Foundation}. This pipeline for spectroscopic data combines (for each raw exposure) the two CCD images from the detector (correcting for geometric distortions) and performs bias subtraction. Although it can perform the wavelength calibration and can correct the curvature across the spatial direction in 2D images, we noticed that it was not perfect. For this reason we opted to perform a manual approach and extracted a small cut in the image around each slit. We performed the wavelength calibration individually for each of these images (slits) and tilt was corrected when necessary.   
The science and sky spectra were extracted (in 1D), and followed by flux calibration. We used \textsc{IRAF} to extract the long-slit spectra for standard stars, and then the routine \texttt{standard} and \texttt{sensfunc} to obtain the  sensitivity curve. This was applied through the \texttt{calibrated} routine to the science spectrum.

For the FORS2 data, we used the FORS2 pipeline v5.5.7 under the EsoReflex environment \citep{esoreflex}. This resulted in flux-calibrated, sky-subtracted 1D spectra for each slit on the mask. However, for some slits the pipeline did not produce suitable spectra, due to multiple objects in the slit, strongly variable nebular emission, slit overlap, and/or strong vignetting at the top of the CCD. For this reason, we also performed the reduction without sky subtraction and manually selected the object and sky extraction regions from the 2D spectrum. For each slit, the automatically and manually extracted spectra were visually inspected, and the best reduction was chosen.

\subsection{Spectral Classification}
\label{s:spectral_classification}

The resolution and wavelength range (as described in the previous section) provide access to a number of spectral features, such as H$\alpha$ (a mass loss tracer for high $\dot{M}$ stars), the TiO bands (present in cool stars), He~\textsc{i} and He~\textsc{ii} lines (indicative of hot stars), various metal lines (notably Fe lines), and the Ca triplet (luminosity indicator). Therefore, we were able to effectively classify the vast majority of our targets. 

Both B[e]SGs and LBVs are characterized by strong emission lines, indicative of their complex CSEs. H$\alpha$ is usually found in very strong emissions and is significantly broadened in the presence of strong stellar winds and/or the presence of a (detached) disk (e.g., \cite{Kraus2019a, Maravelias2018}). There were a number of  He~\textsc{i} lines (at $\lambda\lambda5876.6, 6678.2, 7065.2, 7281.4$) within our observed range, which manifest in the hottest sources. In the quiescence state of LBVs, the presence of He lines indicates hotter sources (of B/A spectral type, which can be observed even with P-Cygni profiles when stellar winds are strong, such as, for example, in \cite{Humphreys2017}). However, when an outburst is triggered and evolves outwards, the temperature temporarily decreases until the ejecta become optically thin. As a result of this temperature shift, the spectral lines typical for the quiescent LBV weaken and metal emission lines strengthen (e.g.,~\cite{Ritchie2009}). During this phase, and depending on the temperature and density conditions of the circumstellar material, they may also display some forbidden Fe lines. B[e]SGs display additional forbidden emission lines, due to their more complex CSEs, with typical examples being [O~\textsc{i}] $\lambda\lambda5577, 6300, 6364$ and [Ca~\textsc{ii}] $\lambda 7291, 7324$. The latter is more evident in the more luminous sources (e.g., \citep{Aret2012, Humphreys2017}).

Therefore, among all sources identified with strong H$\alpha$ emissions, we classified as being B[e]SGs those with evident [O~\textsc{i}] $\lambda6300$ \citep{Lamers1998, Humphreys2017}, and as being LBVs those without. Both classes may display forbidden emission lines from Fe and Ca (e.g., \citep{Aret2016, Condori2019, Humphreys2017}), while all of them display Fe emission lines. We note here that these LBVs are candidate sources, since there is no absolute way to characterize an LBV from a single-epoch spectrum (in contrast to B[e]SGs). It has to be supplemented with more spectroscopic or photometric observations that reveal variability (and possibly the return to a hotter state). We also note that our sample contained more interesting sources that displayed H$\alpha$ in emission (i.e., main sequence O-stars and blue supergiants), but these were left for future papers (e.g., Munoz-Sanchez et al. 2023, IAU S361, submission).

\section{Results}\label{sec3}

\subsection{Statistics}
\label{s:statistics}

From our large observational campaign, we were able to robustly classify (after careful visual inspection) 465 objects in the 12 targeted galaxies (see Table \ref{t:galaxies}). Only 11 out of all of these ($\sim$3\%) contained features in their optical spectra that indicated a B[e] SG/LBV nature (which was the subject of the current work, with the rest being left for future papers). Other stellar sources related to massive stars included mainly RSGs ($\sim$37\%), other Blue Supergiants ($\sim$7\%), and Yellow Supergiants ($\sim5\%$). There was a small number of emission objects ($\sim$2\%), carbon stars ($\sim$6\%), and AGN/QSO and other background galaxies ($\sim$4\%), while another bulk of sources were classified as H~\textsc{ii} regions ($\sim$22\%) and foreground sources ($\sim$14\%). In Table \ref{t:sources} we present the identified objects. We note that, although the same approach was followed for all 12 galaxies, we obtained null results for five of them: IC 10, NGC 1313, Sextans A, M83, NGC 6822. In addition, there were only four objects ($\sim$36\%) with previous spectral information, for which we confirmed or updated classification. It is also interesting to note that $\sim$64\% of these sources were considered priority targets in our survey (Table~\ref{t:sources}, col. 4), while the rest failed to pass our selection criteria (see Section \ref{s:target_selection}). We further discuss these facts in Section \ref{s:discussion}.

\begin{table}[H]

% \centering
\caption{Properties of the sources identified in this work: source ID in this work (column 1), sky coordinates (columns 2 and 3), priority target (column 4), source ID in \textit{Spitzer} (base) catalog (column 5), SNR (column 6), spectral type from this work and literature (columns 7 and 8), and radial velocity from this work (RV,
column 9). } % GM more complete description
\label{t:sources}
%\newcolumntype{C}{>{\centering\arraybackslash}X}

\begin{adjustwidth}{-\extralength}{0cm}
%\centering %% If there is a figure in wide page, please release command \centering

\setlength{\cellWidtha}{\fulllength/9-2\tabcolsep-0in}
\setlength{\cellWidthb}{\fulllength/9-2\tabcolsep+0.1in}
\setlength{\cellWidthc}{\fulllength/9-2\tabcolsep+0.1in}
\setlength{\cellWidthd}{\fulllength/9-2\tabcolsep-0.3in}
\setlength{\cellWidthe}{\fulllength/9-2\tabcolsep-0.2in}
\setlength{\cellWidthf}{\fulllength/9-2\tabcolsep-0in}
\setlength{\cellWidthg}{\fulllength/9-2\tabcolsep-0in}
\setlength{\cellWidthh}{\fulllength/9-2\tabcolsep+0.3in}
\setlength{\cellWidthi}{\fulllength/9-2\tabcolsep+0in}
\scalebox{1}[1]{\begin{tabularx}{\fulllength}{>{\PreserveBackslash\raggedright}m{\cellWidtha}>{\PreserveBackslash\raggedright}m{\cellWidthb}>{\PreserveBackslash\raggedright}m{\cellWidthc}>{\PreserveBackslash\centering}m{\cellWidthd}>{\PreserveBackslash\raggedleft}m{\cellWidthe}>{\PreserveBackslash\raggedleft}m{\cellWidthf}>{\PreserveBackslash\centering}m{\cellWidthg}>{\PreserveBackslash\centering}m{\cellWidthh}>{\PreserveBackslash\raggedleft}m{\cellWidthi}}
\toprule
\textbf{Name} & \textbf{RA} & \textbf{Dec} & \textbf{Prio.} & \textbf{ID~}\boldmath{$^1$} & \textbf{SNR~}\boldmath{$^2$} & \textbf{SpT} & \textbf{Prev. SpT} & \textbf{RV} \\
& \textbf{(J2000)} & \textbf{(J2000)} & & & & & & \textbf{(km s\boldmath{$^{-1}$})} \\
\textbf{(1)} & \textbf{(2)} &\textbf{ (3)} &\textbf{ (4)} & \textbf{(5)} &\textbf{ (6)} & \textbf{(7) }& \textbf{(8) }&\textbf{ (9)} \\

\midrule
WLM-1 & 00:02:02.32 & $-$15:27:43.81  & Y & 95   & 30 & B[e]SG & Fe star {\cite{Britavskiy2015}} & $-$48 $\pm$ 10\\
NGC55-1 & 00:15:09.31 &  $-$39:12:41.62  & Y & 178  & 18 & B[e]SG & -- & 156 $\pm$ 31 \\
NGC55-2 & 00:15:18.54 &  $-$39:13:12.32  & N & 736  & 46 & LBVc & LBVc/WN11 {\cite{Castro2008}} & 105 $\pm$ 38 \\
NGC55-3 & 00:15:37.66 &  $-$39:13:48.68  & N & 2924 & 50 & LBVc & LBVc/WN11 {\cite{Castro2008}} & 202 $\pm$ 30 \\
NGC247-1 & 00:47:02.17 &  $-$20:47:40.13 & Y & 246  & 26 & B[e]SG & B[e]SG {\cite{Solovyeva2020}}  & 217 $\pm$ 12  \\
NGC247-2 & 00:47:03.91 &  $-$20:43:17.22 & N & 1192 & 44 & LBVc & -- & 114 $\pm$ 41 \\
NGC253-1 & 00:47:04.90 &  $-$25:20:44.12 & Y & 739  & 3  & B[e]SG  & -- & 283 $\pm$ 56  \\
NGC300-1 & 00:55:27.93 &  $-$37:44:19.61 & Y & 67   & 44 & B[e]SG & -- & 58 $\pm$ 27 \\ 
NGC300-2 & 00:55:19.17 &  $-$37:40:56.53 & Y & 389  & 9  & B[e]SG & -- & 121 $\pm$ 34  \\ 
NGC3109-1 & 10:03:02.11 &  $-$26:08:58.06 & Y & 188  & 70 & LBVc & -- & 371 $\pm$ 29 \\
NGC7793-1 & 23:57:43.28 &  $-$32:34:01.81 & N & 111  & 19 & B[e]SG c  & -- & 317 $\pm$ 32 \\

\bottomrule
\end{tabularx}}
\end{adjustwidth}
\noindent{\footnotesize{$^1$ This ID corresponds to the \textit{Spitzer} source numbering, as used throughout the ASSESS project  (see Tramper et al., in prep., and Munoz-Sanchez et al., in prep., for the use with full catalogs). 
 $^2$ Estimated by averaging the SNR over the ranges 6000--6150\AA\, and 6950--7100\AA.}}

\end{table}

\subsection{Spectra}
\label{s:spectra}

\begin{figure}[H]
%\begin{center}
   \includegraphics[width=\columnwidth]{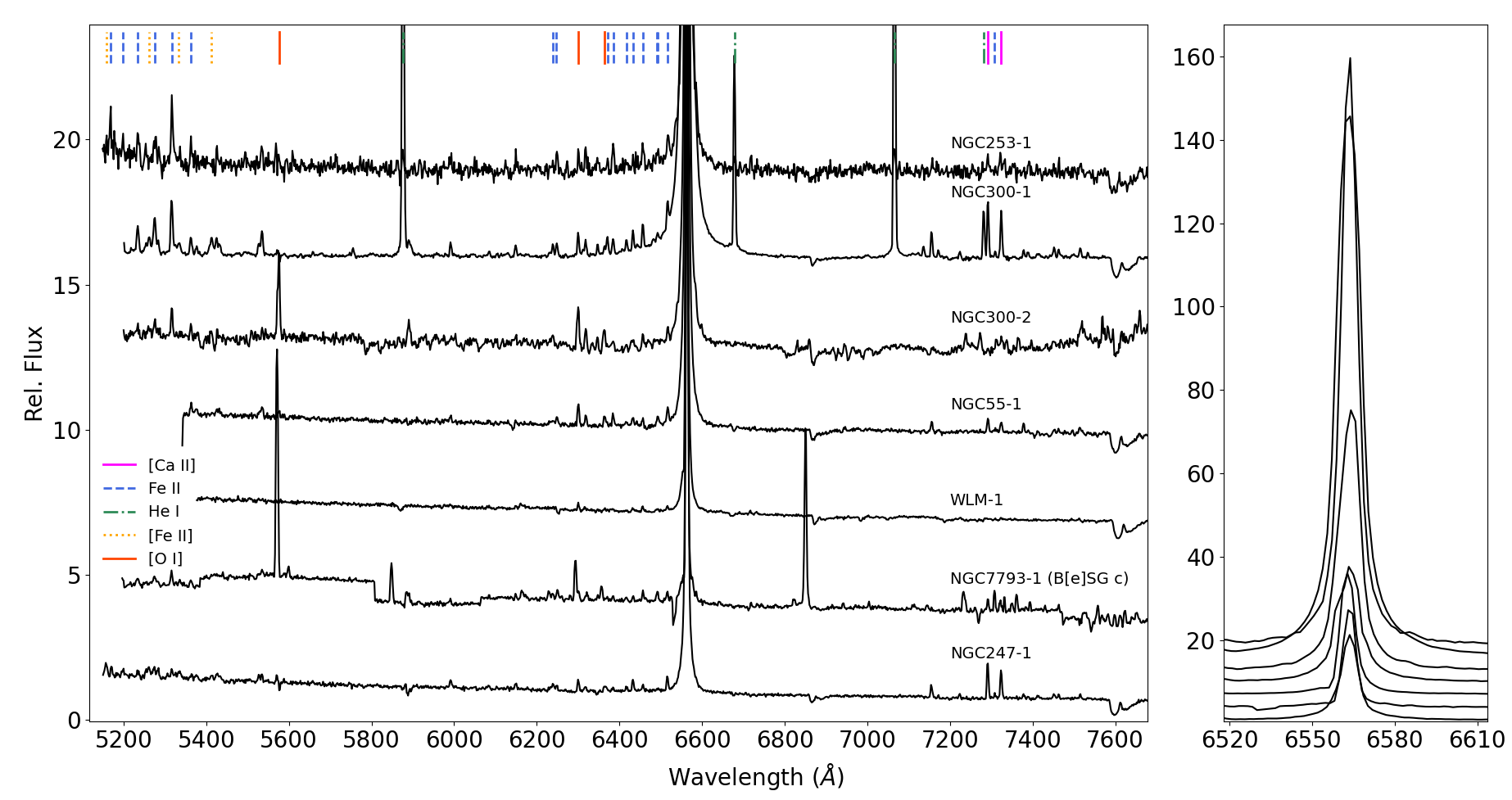}
    \caption{Spectra of objects classified as B[e]SGs (including the B[e]SG candidate NGC7793-1). (Left) The full spectra for all stars with small offsets for better illustration purposes. The most prominent emission features are indicated. (Right) The region around H$\alpha$ is highlighted to emphasize the relative strength of the emission compared to the continuum.}
    \label{f:BeSGs_main}
%\end{center}
\end{figure}

\begin{figure}[H]
%\begin{center}
   \includegraphics[width=\columnwidth]{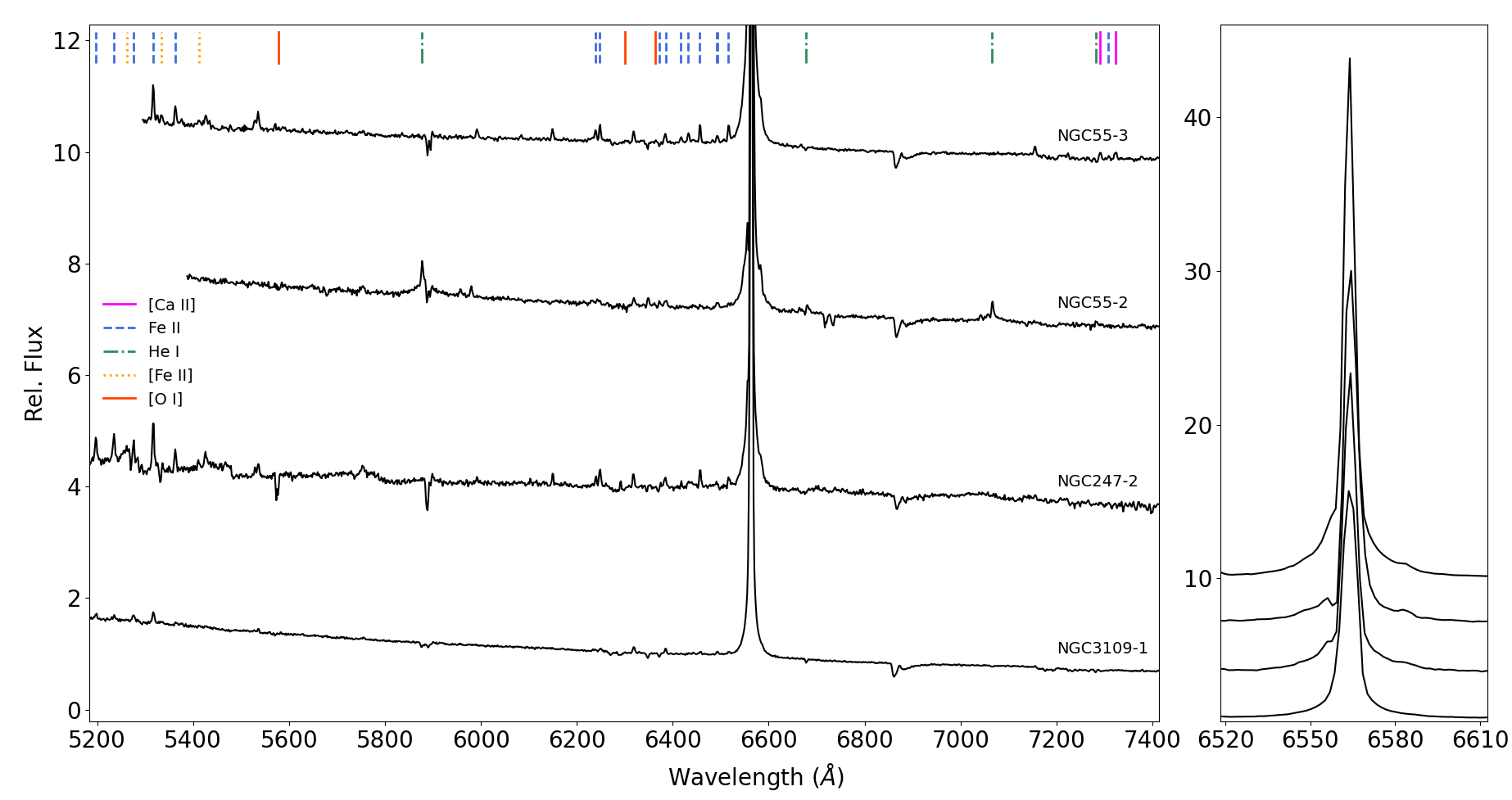}
    \caption{Similar to Figure~\ref{f:BeSGs_main}, but for LBVc. We note the lack of forbidden emission lines. }
    \label{f:LBVs_main}
%\end{center}
\end{figure}

All spectra showed a strong, broadened H$\alpha$ component, accompanied by several other characteristic emission lines. We present their spectra in Figures \ref{f:BeSGs_main} and \ref{f:LBVs_main}, where the strength of the H$\alpha$ emission for all objects is highlighted in the right panel. The order of the spectra (from top to bottom) was one of decreasing H$\alpha$ strength.  

\begin{figure}[H]
%\begin{center}
    \includegraphics[width=\columnwidth]{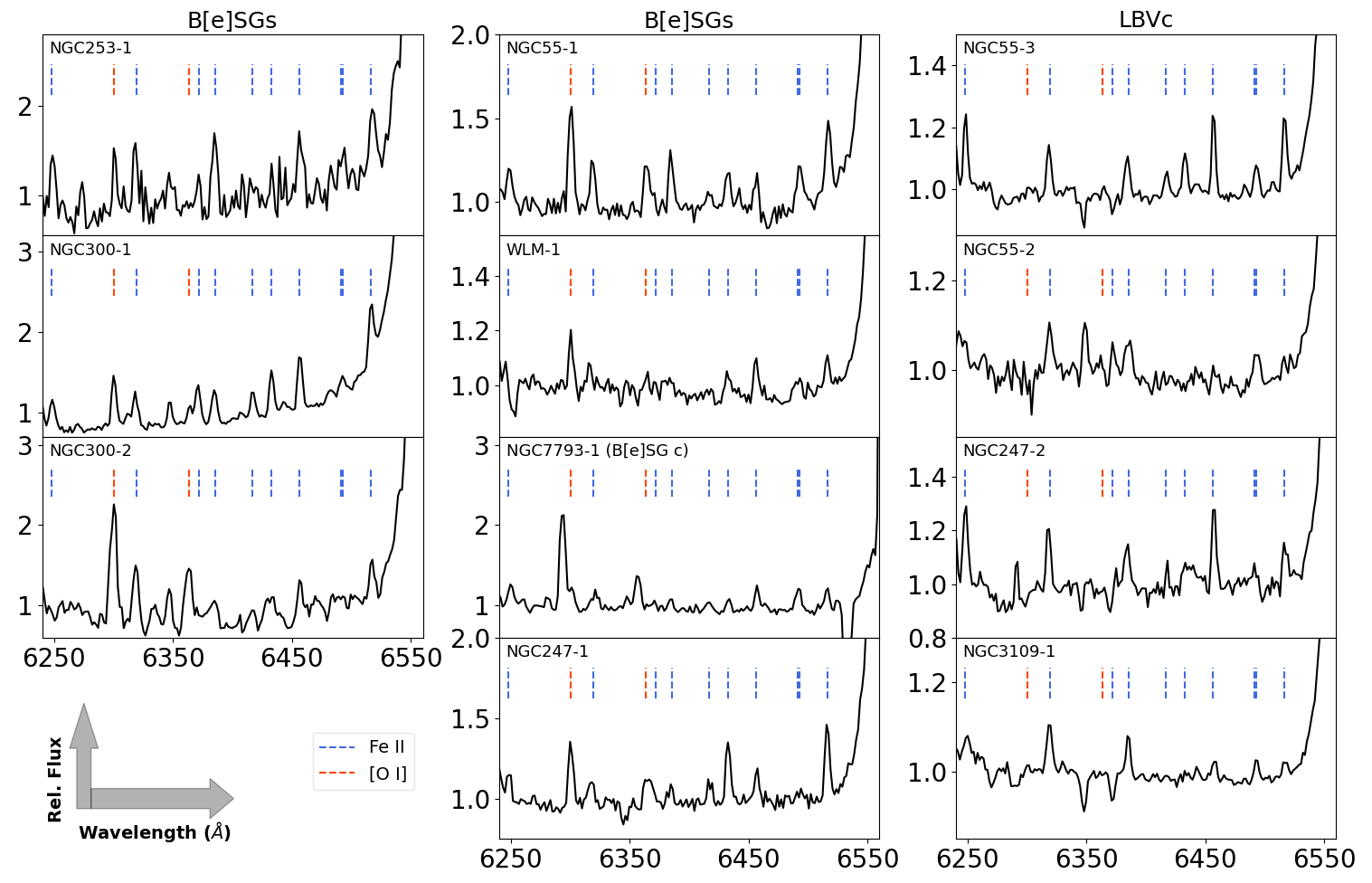}
    \caption{The region between the [O \textsc{i}] and H$\alpha$ line, that showcases multiple Fe \textsc{ii} emission lines. We note the clear presence of [O \textsc{i}] $\lambda$6300 line for the B[e]SGs (left and middle panels, with the exception of the candidate NGC7793-1, due to the problematic spectrum; see text for more) and its absence from the LBVc spectra (right panel).  
    }
    \label{f:Fe_lines}
%\end{center}
\end{figure}

We identified a series of Fe~\textsc{ii} emission lines in the left wing of H$\alpha$ ($\sim$6200--6500~\AA), and, when the spectrum extended far enough to bluer wavelengths, we identified another series ranging from roughly $\sim$5100--5400~\AA. Figure \ref{f:Fe_lines} showcases these lines in a zoom-in on the $\sim$6200--6500~\AA\, region. We used the Fe~\textsc{ii} emission lines in this region to correct for the radial velocity (RV) shift. The obtained RV values are shown in column 9 of Table \ref{t:sources}. Therefore, we verified that the RVs were in agreement with the motion of their host galaxies, confirming that these stars were, indeed, of extragalactic origin. 

\begin{figure}[H]
%\begin{center}
  \includegraphics[width=\columnwidth]{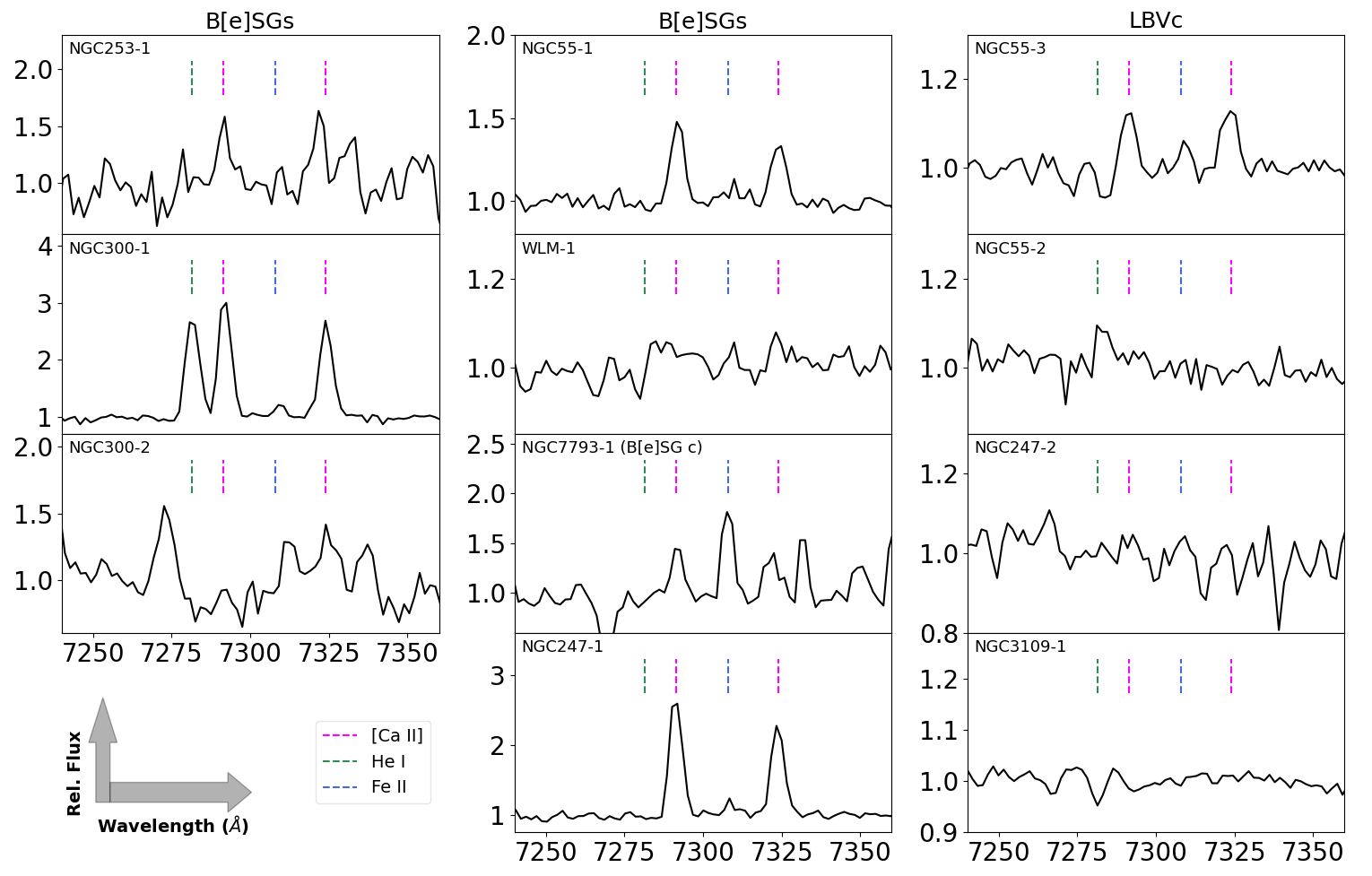}
    \caption{The region around the [Ca \textsc{ii}] emission doublet. Its presence is evident in some B[e]SGs (left and middle panels, including NGC7793-1 candidate source, that suffers from data reduction artifacts due to slit overlaps), while LBVc (right panel) do not typically exhibit these lines (except for NGC55-3).}
    \label{f:Ca_lines}
%\end{center}
\end{figure}

According to the classification criteria presented in Section \ref{s:spectral_classification}, we robustly identified 6~sources as being B[e]SGs: WLM-1, NGC55-1, NGC247-1, NGC253-1, NGC300-1, and NGC300-2. Figure~\ref{f:BeSGs_main} presents the full spectra for the B[e]SGs, while Figure~\ref{f:Fe_lines} shows the characteristic [O~\textsc{i}] $\lambda6300$ line. It is particularly interesting to note the very strong He~\textsc{i} lines of NGC300-1. These emission lines require a hotter formation region, such as a spherical or a bipolar shell formed by a strong stellar wind, in addition to the structures that give rise to the forbidden emission features. We also note the absence of [Fe~\textsc{ii}] lines for the WLM-1, NGC253-1, and NGC300-2 sources. Half of the sources (NGC55-1, NGC247-1, NGC300-1) displayed strong [Ca~\textsc{ii}] emission lines, while for one source (NGC253-1) they were very faint (limited by the noise), and were totally absent for two of the sources (WLM-1, NGC300-2; see Figure~\ref{f:Ca_lines}). These lines were stronger in luminous sources (e.g., \citep{Aret2016, Condori2019}). The very low SNR for the NGC253-1 and NGC300-2 (see Table~\ref{t:sources}, column 6) justified the lack of Fe and Ca lines. In the case of WLM-1, the SNR was sufficiently good that the lack of forbidden Fe lines should be considered a real non-detection (similar to source WLM 23 from \cite{Britavskiy2015}. We further discuss this in Section \ref{s:previous_classification}).  Unfortunately, due to overlapping slits in the mask design, some of these spectra suffered from artifacts from the reduction processing (in particular, NGC300-2). Although the B[e] phenomenon can also characterize other types of objects, we noticed a lack of dominant emission lines, such as nebular lines ([N~\textsc{II}]~$\lambda\lambda6548, 6583$, [S ~\textsc{ii}]~$\lambda\lambda6717,6731$, [Ar~\textsc{iii}]~$\lambda7135$), present in  planetary nebulae (e.g., \cite{Stasinska2013, Ilkiewicz2017}), O~\textsc{VI} Raman-scattered lines ($\lambda\lambda6830,7088$) of symbiotic systems (e.g., \cite{Ilkiewicz2017, Akras2021}), or even the absorption lines of Li~\textsc{i}~$6708$ present in young stellar objects (e.g., \cite{Megeath2022}). Moreover, during the visual screening of all these spectra, objects with such characteristic lines would be classified differently, as all possible objects were considered. Additionally, at the distances we were looking at, we were mainly probing the upper part of the Hertzsprung--Russell diagram, while their RVs were relatively compatible (within their error margins) with those of their host galaxies. Our \textit{Gaia} cleaning approach removed the majority of the foreground sources (naturally, a small fraction remained hidden in our target lists). Therefore, we consider these objects to be strong supergiant candidates.

We characterized as LBVc the following 4 sources: NGC55-2, NGC55-3, NGC247-2, and  NGC3109-1 (see Figure~\ref{f:LBVs_main}). NGC55-2 was the hotter of all these sources as it was the only LBVc with all He~\textsc{i} lines in emission. NGC3109-1 displayed He~\textsc{i} lines in absorption, while the rest did not show any of these lines. During the outbursts the He~\textsc{i} lines decrease and vanish, as the temperature and the density (due to the expanding pseudo-photosphere) drop significantly to allow for other lines to form. It is during these cooler states that Fe lines become evident in LBVs. Depending on the conditions, forbidden emission lines may form. This was the case with NGC55-3, which displayed the [Ca~\textsc{ii}] lines in emission, along with a few [Fe~\textsc{ii}] lines. The other sources did not show any forbidden lines. Similar to the B[e]SG spectra, there were unavoidable residuals and artifacts, due to the slit overlap and reduction issues. 

Of these cases, NGC7793-1 was the most extreme example\endnote{Features at $\lambda\lambda \sim$5577, 5811 (step), 5846, 6855, and the region around the [Ca~\textsc{ii}] lines.}. The region at [O~\textsc{i}] $\lambda6300$ was highly contaminated with a sky residual line from another source in the slit. Therefore, we could not conclude whether this line existed or not. We noticed the presence of some [Fe~\textsc{ii}] and the [Ca~\textsc{ii}] lines, but a B[e]SG or LBV classification solely from this spectrum was not possible. However, additional information could be retrieved from photometry (see Section~\ref{s:photometry}), so that we could propose a B[e]SG candidate (B[e]SG c) classification for NGC7793-1. 

The final classification for each star is provided in column 7 of Table \ref{t:sources}.

\vspace{-9pt}

\vspace{-9pt}

\newpage
\subsection{Light Curves and Variability}
\label{s:light_variability}

We collected variability information for all targets from both Pan-STARRS DR2\endnote{\highlighting{\url{https://catalogs.mast.stsci.edu/panstarrs/}}} and the VizieR\endnote{\highlighting{\url{http://vizier.cds.unistra.fr/}}} services. We found four sources (WLM-1, NGC247-1, NGC274-2, and NGC3109-1) with data in the Pan-STARRS DR2 release\endnote{There were only a couple of detections (epochs) for NGC253-1, which did not provide any meaningful information, and, therefore, we did not consider them. A declination of about $-25^\circ$ was very close to the limit of the survey. All other galaxies with southern declination than $-30^\circ$, i.e., NGC 55, NGC 300, and NGC 7793, were not visible.} (with an approximate coverage between 2010 and 2014). We considered only values with \texttt{psfQfPerfect}>0.9 to select the best data. For three sources (NGC55-2, NGC55-3, and NGC247-1) we found additional data in the catalog of large-amplitude variables from \textit{Gaia} DR2 (covering 2014 to 2016; \cite{Mowlavi2021}), and NGC3109-1 had already been reported as a variable \cite{Menzies2019}. In Table~\ref{t:variability} we summarize the collected information for all sources and their corresponding magnitude differences (peak-to-peak) for all (5) Pan-STARRS filters, the two \textit{Gaia} filters (for which we doubled the quoted values in the catalog to match the Pan-STARRS definition of magnitude difference, and some additional variability studies).  

\begin{figure}[H]
\begin{adjustwidth}{-\extralength}{0cm}
\centering
    \includegraphics[width=0.65\columnwidth]{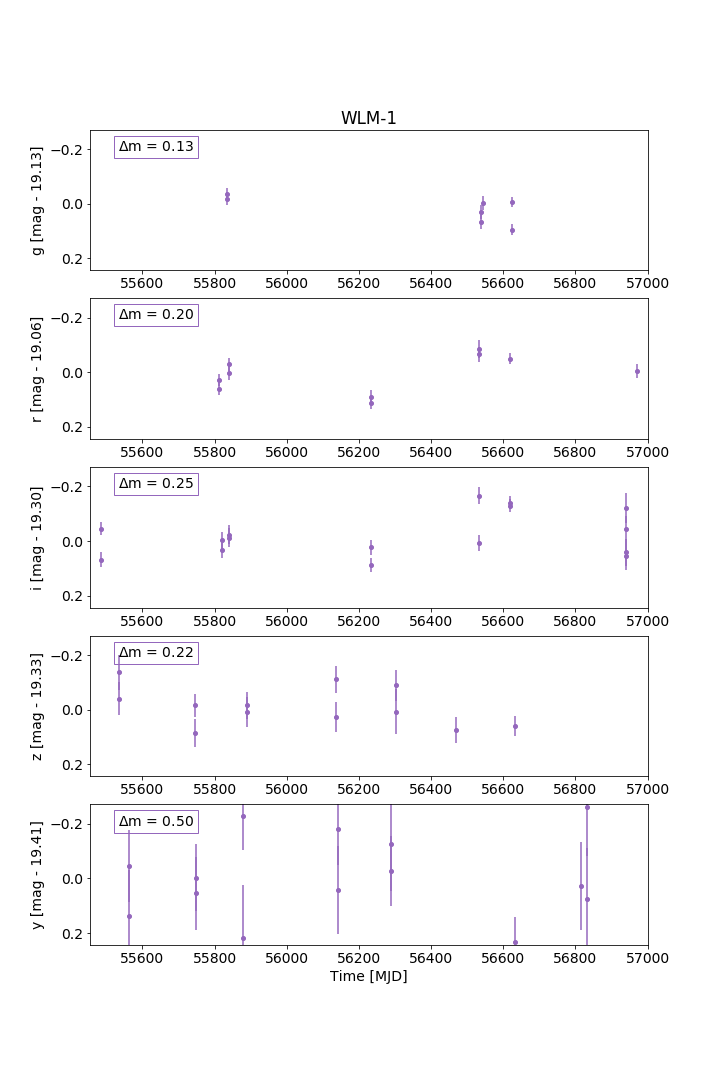}
    \includegraphics[width=0.65\columnwidth]{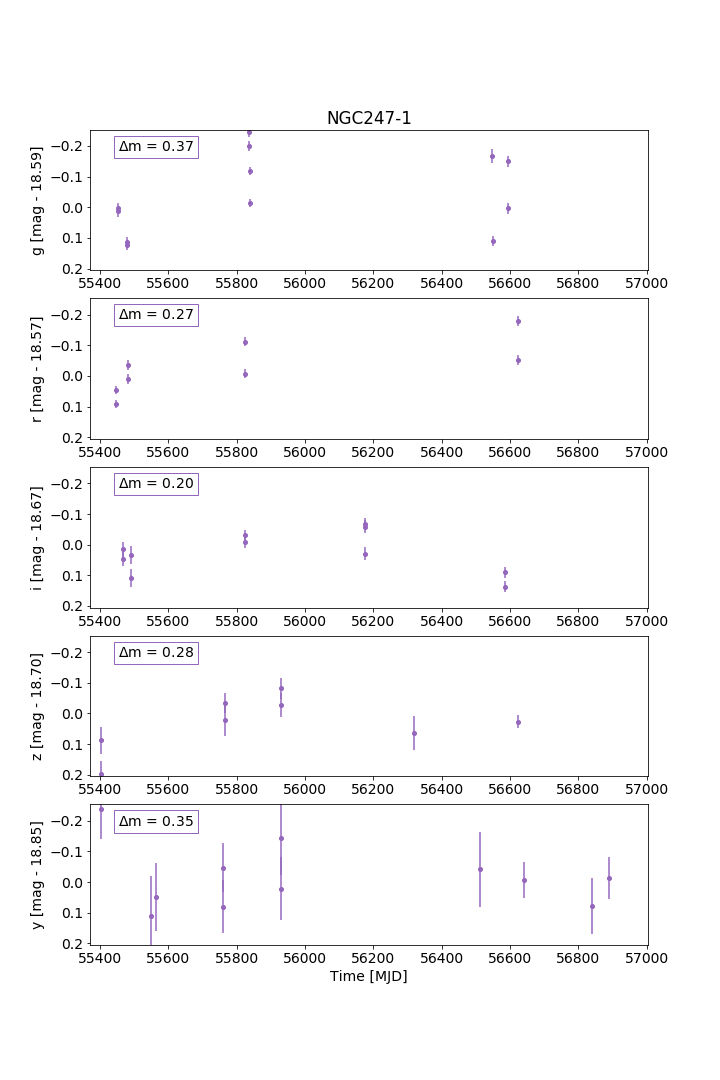}
    \end{adjustwidth}
    \caption{The light curves from the Pan-STARRS survey for the B[e]SGs WLM-1 and NGC247-1. Each panel (per filter) shows the difference of each epoch from the mean value (noted on the y-axis label). See text for more.}
    \label{f:B[e]SGs_lcs}

 \end{figure}

\begin{figure}[H]
\begin{adjustwidth}{-\extralength}{0cm}
\centering
    \includegraphics[width=0.65\columnwidth]{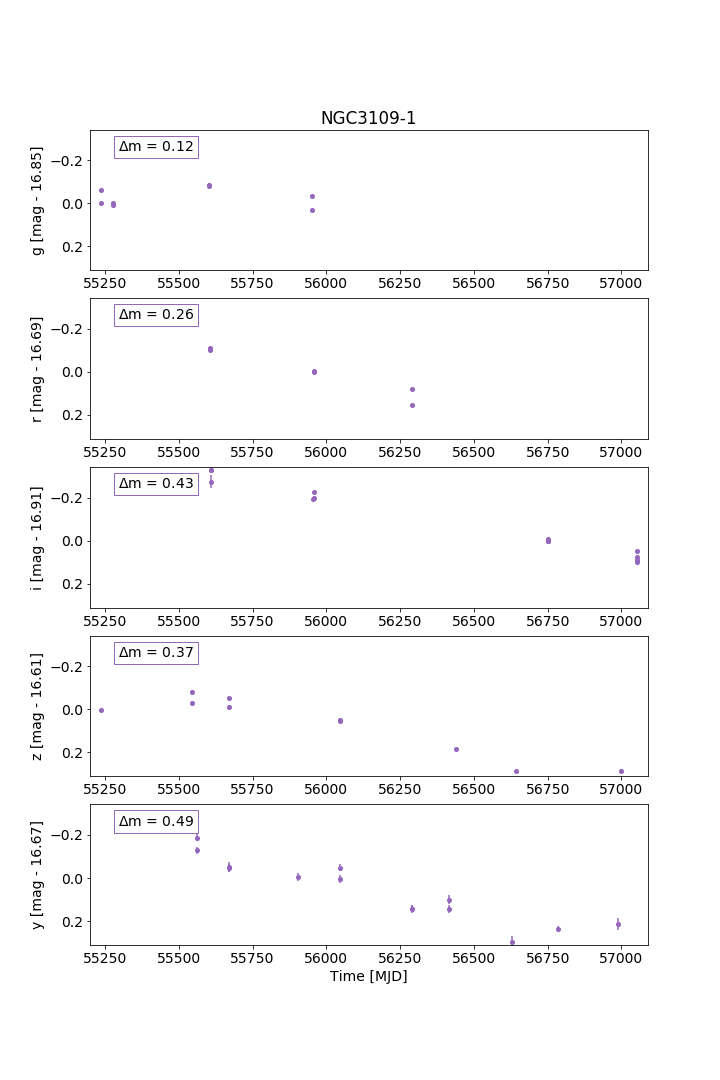}
    \includegraphics[width=0.65\columnwidth]{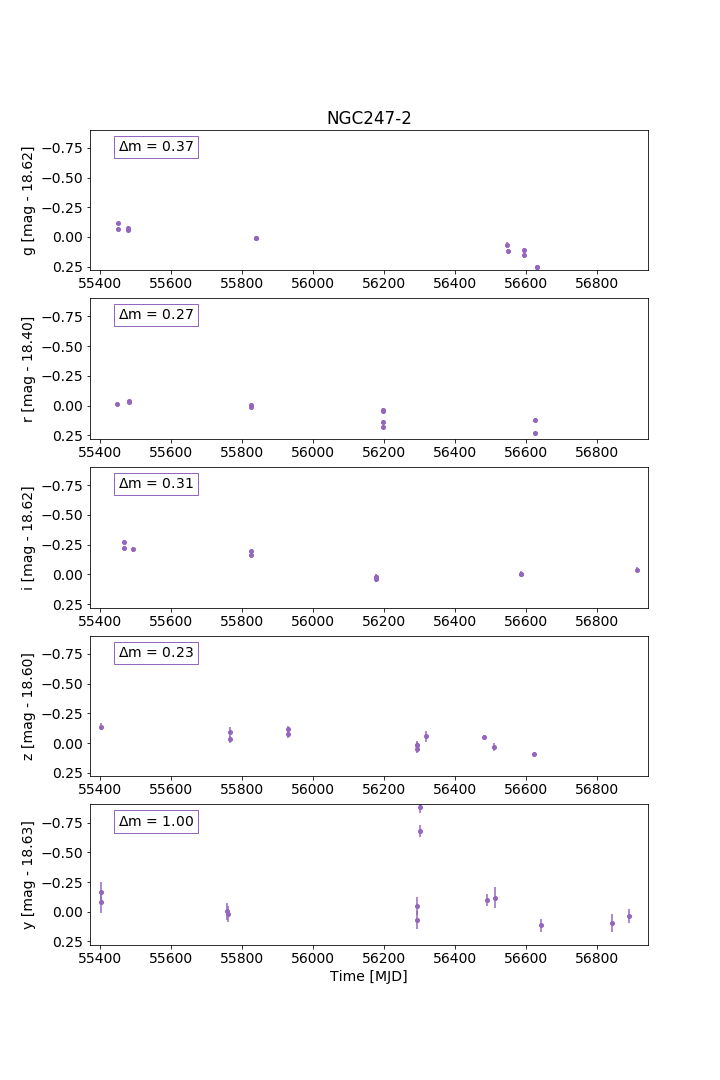}
  \end{adjustwidth}   
    \caption{Same as Figure~\ref{f:B[e]SGs_lcs}, but for the candidate LBVs NGC3109-1 and NGC247-2. }
    \label{f:LBVs_lcs}

 \end{figure}

In total, we found light curves for two B[e]SGs (WLM-1 and NGC247-1) and four LBVc (NGC55-2, NGC55-3, NGC247-1, and NGC3109-1). We show the Pan-STARRS light curves in Figure~\ref{f:B[e]SGs_lcs} and \ref{f:LBVs_lcs}, where we plot the magnitude difference at each epoch with the mean for the particular filter (indicated on the y-axis label). For the B[e]SGs we noticed a (mean) variability of 0.25-0.3 mag, while for the LBVs it was slightly larger, at 0.3--0.44 mag. There were no obvious trends in the B[e]SG light curves, while, in the case of NGC3109-1, a dimming across all filters was observed. \citet{Menzies2019} also detected such a trend, although smaller, for this target, due to the different filters used. Limited by the photometric data, they argued that a background galaxy or AGN was not excluded, but, given our spectrum and its consistent RV value with its host galaxy, we could actually verify its stellar nature. For NGC247-2, the light curves were generally flatter. There was a noticeable peak present in the $y$ light curve (at MJD$\sim56300$ days), which was not evident in the other filters (although we note  that there were no observations around the same epoch). The quality flags corresponding to these particular points did not show any issue. However, we should be cautious with this, as further mining of the data is needed to reveal if this is a real event or an artifact. 

NGC247-1 was the only source for which we had multiple sources of variability information. Very good agreement between the Pan-STARRS and \textit{Gaia} data is evident, and consistent with the value quoted by \citet{Solovyeva2020} ($\Delta$V=0.29 $\pm$ 0.09~mag). Although \citet{Davidge2021} quoted a smaller value ($\Delta g' \sim 0.1$ mag), their time coverage was limited to about 6 months, a time frame that definitely does not cover the whole variability cycles for these sources.

Traditionally, LBVs are considered variable at many scales (e.g., \cite{vanGenderen2001, Martin2017, Davidson2020, Weis2020}). {{The (optical) S Dor variability is} of the order of 0.1 mag to about 2.5 mag with cycles ranging from years to decades. The giant eruptions, although much more energetic ($\sim$5 mag) are less frequent events (a time frame in the order of centuries), and, therefore, a smaller subgroup of LBVs have been observed to display such events.} On the other hand, the B[e]SGs are considered more stable, with variability that does not exceed $\sim$0.2 mag (optical; \cite{Lamers1998}). However, this is changing and significant variability is observed, due to binary interactions and possible pulsations (e.g., \cite{Kraus2016, Maravelias2018, Porter2021}). Therefore, it is not surprising to observe similar magnitude differences between the two classes.

\begin{table} [H]
\caption{Variability information for our sample (source IDs, column 1), as provided by Pan-STARRS DR2 data (columns 2-6), \textit{Gaia} (columns 7 and 8), and literature (column 9).} 
 \label{t:variability}
%\vspace{-9pt}
%\newcolumntype{C}{>{\centering\arraybackslash}X}
\setlength{\tabcolsep}{1.8mm}
\resizebox{\textwidth}{!}{
\begin{tabularx}{\textwidth}{lcccccccc}
\toprule
\textbf{Name} & \multicolumn{5}{c}{\textbf{Pan-STARRS DR2}} & \multicolumn{2}{c}{\textbf{\textit{Gaia} DR2 {\cite{Mowlavi2021}}}} & \textbf{Other} \\
& \boldmath{$\Delta\,g$} & \boldmath{$\Delta\,r$} & \boldmath{$\Delta\,i$} & \boldmath{$\Delta\,z$} & \boldmath{$\Delta\,y$} & \boldmath{$\Delta$}\textbf{BP} & \boldmath{$\Delta$}\textbf{RP} & \\
 & \textbf{(mag)}  & \textbf{(mag) } & \textbf{(mag)}  & \textbf{(mag)}  & \textbf{(mag)}  & \textbf{(mag)}  &\textbf{ (mag)} \\
 \textbf{(1) }&\textbf{ (2)} & \textbf{(3)} &\textbf{ (4)} & \textbf{(5) }& \textbf{(6)} & \textbf{(7)} & \textbf{(8)} & \textbf{(9)} \\
\midrule

WLM-1 & 0.13 & 0.20 & 0.25 & 0.22 & 0.50 & -- & -- & -- \\
NGC55-1 & -- & -- & -- & -- & -- & -- & -- & --  \\
NGC55-2 & -- & -- & -- & -- & -- & 0.24 & 0.18 & --  \\
NGC55-3 & -- & -- & -- & -- & -- & 0.24 & 0.18 & --  \\
NGC247-1 & 0.37 & 0.27 & 0.20 & 0.28 & 0.35 & 0.36 & 0.18 &  $\Delta$V = 0.29 $\pm$ 0.09 {\cite{Solovyeva2020}} \\
 &  &  &  &  &  &  &  & $\Delta g' \sim 0.1$ {\cite{Davidge2021}}  \\
NGC247-2 & 0.37 & 0.27 & 0.31 & 0.23 & 1.00 & -- & -- & -- \\
NGC253-1 * & -- & -- & -- & -- & -- & -- & -- & -- \\
NGC300-1 & -- & -- & -- & -- & -- & -- & -- & --  \\
NGC300-2 & -- & -- & -- & -- & -- & -- & -- & --  \\
NGC3109-1 & 0.12 & 0.26 & 0.43 & 0.37 & 0.49 & -- & -- &  $\Delta J=0.08$ {\cite{Menzies2019}}  \\ %[MWF2019] 1060
NGC7793-1 & -- & -- & -- & -- & -- & -- & -- & --  \\

\bottomrule
\end{tabularx}}

\noindent\footnotesize{~* Only three epochs of observations, so not considered.  }

\end{table}

\vspace{-14pt}

\section{Discussion}%\label{sec4}
\label{s:discussion}

\subsection{Demographics}

As mentioned already in Section \ref{s:statistics}, we did not detect B[e]SGs or LBVs in the following five (out of 12) galaxies: IC 10, NGC 1313 Sextans A, M83, NGC 6822. 

M83 and NGC 1313 are the most distant  galaxies (at 4.9 and 4.6 Mpc, respectively) and confusion becomes an important issue (unsurprisingly, M83 is the galaxy for which we detected the most H~\textsc{ii} regions; see Tramper et al., in prep.). Due to the spatial resolution of \textit{Spitzer} and the increasing distance of some of our target galaxies, H~\textsc{ii} regions or other point-like objects (e.g., clusters) were included in the point-source catalogs and, therefore, considered to be viable targets in our priority system. The farthest galaxies, for which the majority of observed targets were, indeed, resolved point sources and at least one was either an LBVc or a B[e]SG, were NGC 7793 and NGC 253 (at $\sim$3.4~Mpc). Therefore, the null detections for IC 10 and NGC 6822 (less than 1 Mpc) and for Sextans A (at 1.34 Mpc) were not due to distance and confusion. 

\citet{Massey2007} detected one LBV in NGC 6822 (J194503.77-145619.1) and three in IC 10 (J002012.13+591848.0, J002016.48+591906.9,  J002020.35+591837.6). Our inability to recover these targets was due to two reasons. Firstly, we imposed strict criteria to prioritize our target selection (see Section \ref{s:target_selection}) based on relative strong IR luminosity and color. Almost all of these targets (except for IC 10 J002020.35+591837.6) had m$_{[4.5]}>15.5$ mags, which directly excluded them from further consideration. This was further supported by the fact that four out of or our 11 discoveries initially did not pass as a priority target (see Table~\ref{t:sources}), but were observed as ``filler" stars (see Section \ref{s:target_selection}). This was particularly important for galaxies with smaller sizes, where only one (IC 10, Sextans A) or two pointings (NGC 6822) were performed. Therefore, the second reason was the limitations that arose from the particular pointing(s) to the galaxy, as targets might have been located out of the field-of-view or at a sensor's gap (which was the case for IC 10 J002020.35 + 591837.6), and therefore not be observable. Other reasons (not corresponding to the aforementioned targets) that could impact the selection of a target or render its spectrum useless include overlapping slits, a poor wavelength calibration and/or SNR, or other reduction issues. 

In the case of NGC 55, four LBVs (including two candidates) are known \cite{Castro2008}. Two of them (the candidates) were recovered from our survey (NGC55-2 and NGC55-3 as B\_13 and B\_34, respectively) as LBVc (see Section \ref{s:spectra}). The other two (C1\_30,00:14:59.91,-39:12:11.88 and A\_42,0:16:09.69,-39:16:13.44) were sources outside the region investigated by \citet{Williams2016}, so without any  \textit{Spitzer} data to be included in our base catalogs.

In total, our approach was successful in detecting these populations, and it was mainly limited by technical issues.

\subsection{Comparison with Previous Classifications}
\label{s:previous_classification}

Four of our sources had previous classifications (see Table~\ref{t:sources}). WLM-1 had been identified as an H$\alpha$ source previously \cite{Massey2007}, through a photometric survey, and identified as an Fe line star through spectroscopic observations (WLM 23 in \cite{Britavskiy2015}). Even though the presence of the [O~\textsc{i}] $\lambda6300$ line was noted, the source was not classified as a B[e]SG, due to the lack of forbidden Fe lines (see e.g., \cite{Clark2012, Humphreys2014, Humphreys2017} on Fe stars). Therefore, we updated its classification to a B[e]SG from an Fe star.
We also noted that our spectrum (obtained on November 2020) was very similar to theirs (obtained on December 2012), which might indicate that the star was rather stable over this eight-year period (however, this should be treated with caution due to the lack of systematic observations).

NGC55-2 and NGC55-3 had been identified as candidate LBV/WN11 (ids B\_34 and B\_13, respectively), with both Balmer and He~\textsc{i} lines in emission and with P-Cygni profiles~\cite{Castro2008}. Their spectra were within the 3800--5000~\AA\, range and outside ours. However, given that the diagnostic [O~\textsc{i}] line was not present, we classified both of these sources as LBVc, consistent with the previous results\endnote{As our observations were obtained from different epochs (October--December 2020) than those by \citet{Castro2008} (November 2004) the spectra appearance might have changed, but there was no wavelength overlap to confirm this.}. 

For NGC247-1 we provided a classification of B[e]SG, similar to what was suggested by \citet{Solovyeva2020}. We note here that their spectral coverage was $\sim$4400--7400~\AA\, which overlapped with our observed range. Hence, we can also comment that no significant differences existed between the two observations (October 2018 and December 2020 by \citet{Solovyeva2020} and our observations, respectively), although this time difference is rather small with respect to the variability timescales for these sources \cite{Lamers1998, Kraus2016, Maravelias2018}.
 
Therefore, we confirmed the previous classifications for three out of four sources, leaving us with 6 new B[e]SGs (including the reclassified Fe star and the candidate NGC7793-1) and 2 LBVc. The majority ($\sim$72\%) of our findings are genuine discoveries and, as such, contribute greatly to the pool of extragalactic B[e]SGs, in particular.

\subsection{Separating the Two Classes with Photometry}
\label{s:photometry}

The total numbers of B[e]SGs and LBVs (even including candidates) are definitely small. Combined with the uncertainty pertaining to their roles in stellar evolution theory (e.g., B[e]SGs are not predicted by any code) it is easy to grasp why we really need larger samples and from different galactic environments, to fully understand these sources. Photometric data are typically used to pinpoint interesting candidates. These kinds of diagnostics exist mainly for IR, due to the presence of dust around these objects. 

\begin{figure}[H]
\begin{adjustwidth}{-\extralength}{0cm}
\centering
    \includegraphics[width=0.65\columnwidth]{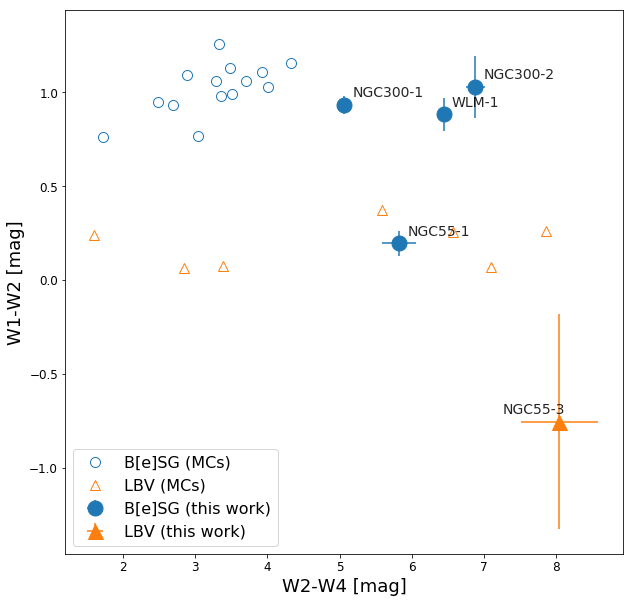}
    \includegraphics[width=0.65\columnwidth]{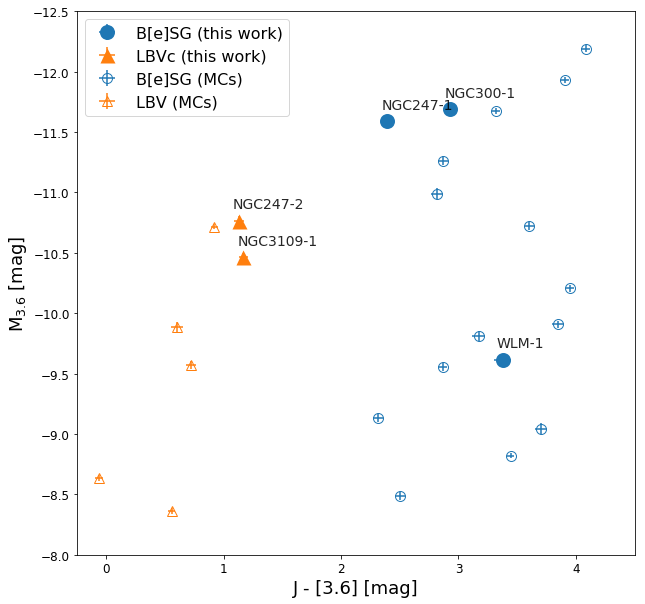}
 \end{adjustwidth}   
    \caption{ (Left) The mid-IR \textit{WISE} CCD for B[e]SGs and LBVs, including sources from the MCs (after~\cite{Kraus2019a}) and our sample (for 5 out of 11 sources with \textit{WISE} data). In general, the separation also holds for the new sources, with the exception of NGC55-1 (see text for more).  (Right) IR CMD combining near-IR \textit{J}-band (available for only five of our sources) with \textit{Spitzer} [3.6]. We notice that, in this case, the newly found sources are consistent with the positions of the MC sources.  
    }  
    \label{f:wise_plot}
\end{figure}

\citet{Bonanos2009} found the B[e]SG, LBVs and RSGs to be among the most luminous sources in the mid-IR, using a color-magnitude diagram (CMD) with a combination of near-IR (2MASS) and mid-IR (\textit{Spitzer}) J-[3.6] and [3.6]-[4.5] for the massive stars in the Large Magellanic Cloud (with a similar work for the Small Magellanic Cloud presented in \cite{Bonanos2010}). In the most recent census of B[e]SGs, \citet{Kraus2019a}  presented color--color diagrams (CCD) to highlight the separation between B[e]SGs and LBVs (see their Figure \ref{f:B[e]SGs_lcs}). Indeed, by using the 2MASS near-IR colors $H-K$ and $J-H$ and mid-IR \textit{WISE} W2-W4 and W1-W2 the two classes are distinct. This is the result of the hot dust component in the B[e]SGs, (formed in the denser disk/ring-like CSE closer to the star) which intensifies the near- and mid-IR excesses, compared to the LBVs (which form dust further away as the wind mass-loss and/or outburst material dissipates). Therefore, the location of a source in these diagrams may be used to verify its nature. We attempted to replicate these aforementioned diagrams by adding the new sources. However, one strong limitation was the lack of data for our sample. For the mid-IR \textit{WISE} \cite{WISE2012} we found data for 5 (out of 11) sources (see Table~\ref{t:photometry}). Using the data for 21 stars (excepting LHA 120-S 111) provided in \cite{Kraus2019a} we plot, in Figure \ref{f:wise_plot}, the \textit{WISE} colors for the MC sources and our 5 objects. We notice that, in general, the newly discovered sources are almost consistent with the loci of the MC sources, with the exception of NGC55-1. The new B[e]SG extend the W2-W4 color further to the red, while the LBVc NGC55-3 extended the W1-W2 color to the blue. Errors were plotted in the cases where they were available\endnote{Only NGC55-1 had an error estimate in the W4 band, while the rest of the sources did not. For all other sources we could only plot W1--W2 errors.}. The errors provided for NGC55-1 were (numerically) small and placed it within the locus of LBV. However, caution should be taken with \textit{WISE} photometry, as the resolution from W1 to W4 worsens significantly, and, combined with the distance of our galaxies, the photometric measurements could be strongly affected by confusion due to crowding (e.g., for both NGC 55 and NGC300 at $\sim$2~Mpc). Combined with the position (and the uncertainty) of the LBVc NGC55-3, we might also be looking at a systematic offset of these populations. Unfortunately, the points in this plot are too scarce to make a robust examination of how the different galactic environments (e.g., metallicity, extinction effects) affect the positions of these populations. 

 \begin{table}[H] 
\caption{Photometry in \textit{Gaia} DR3 (columns 2--7), VHS DR5 (columns 8-13),  \textit{Spitzer} (columns 14--23), and WISE (columns 24--31). \label{t:photometry}}
%\newcolumntype{C}{>{\centering\arraybackslash}X}

\setlength{\tabcolsep}{3.3mm}
\resizebox{\textwidth}{!}{
\begin{tabularx}{\textwidth}{lrrrrrrrr}
\toprule
  \textbf{ID} &
  \textbf{BP} &
  \boldmath{$\sigma_{BP}$} &
  \textbf{RP} &
  \boldmath{$\sigma_{RP}$} &
  \textbf{G} &
  \textbf{...} &
  \textbf{W4} &
  \boldmath{$\sigma_{W4}$} \\
   & \textbf{[mag]} & \textbf{[mag] }& \textbf{[mag]} & 
   \textbf{[mag]} & \textbf{[mag]} &  & \textbf{[mag]} & \textbf{[mag]} \\
   \textbf{(1)} & \textbf{(2)} & \textbf{(3)} & \textbf{(4)} & \textbf{(5)} & \textbf{(6)} & \textbf{...} & \textbf{(30)} & \textbf{(31)} \\
   
  \midrule
WLM-1 & 19.132 & 0.029 & 18.919 & 0.025 & 19.252 & ... & 8.278 & $-$999\\
NGC55-1 & 19.797 & 0.053 & 19.199 & 0.045 & 19.552 & ... & 8.115 & 0.233\\
NGC55-2 & 18.620 & 0.018 & 18.332 & 0.024 & 18.691 & ... & $-$999 & $-$999\\
NGC55-3 & 18.266 & 0.023 & 17.772 & 0.026 & 18.148 & ... & 9.091 & $-$999\\
NGC247-1 & 18.602 & 0.036 & 18.341 & 0.028 & 18.723 & ... & $-$999 & $-$999\\
NGC247-2 & 18.798 & 0.026 & 18.344 & 0.029 & 18.573 & ... & $-$999 & $-$999\\
NGC300-1 & 18.557 & 0.013 & 17.784 & 0.013 & 18.323 & ... & 8.897 & $-$999\\
NGC300-2 & 20.866 & 0.093 & 20.556 & 0.113 & 20.742 & ... & 9.013 & $-$999\\
NGC3109-1 & 17.173 & 0.014 & 16.775 & 0.018 & 17.060 & ... & $-$999 & $-$999\\
NGC7793-1 & 19.635 & 0.043 & 19.460 & 0.053 & 19.520 & ... & $-$999 & $-$999\\
\bottomrule
\end{tabularx}}

\noindent\footnotesize{~Note: The table is available in its entirety at the CDS.}

\end{table}

We were unable to construct the $J\text{-}H$ vs. $H\text{-}K$ CCD because of the lack of 2MASS data for our sources (only for NGC3109-1 did data exist; 2MASS point source catalog; \cite{2MASS2003}), due to the shallowness of the survey and the distances of our target galaxies. However, we were able to acquire $J$ photometry from the VHS DR5 for 5 of our sources (including NGC3109-1;~\cite{VISTA2021}). Equipped with both $J$ and [3.6] photometry we plot, in Figure~\ref{f:wise_plot}, the equivalent CMD plot presented in \cite{Bonanos2009}, where the underlying MC objects were the same as in~\cite{Kraus2019a}. We notice excellent agreement of all new sources to their corresponding classes. 

Once again we were hampered by a lack of data for our sample. We could remedy this using the complete data from \textit{Spitzer} and \textit{Gaia} surveys (missing NGC253-1 from our sample without \textit{Gaia} data). In order to consider the MC sources, we used the \textit{Gaia} DR3 \cite{Gaia2016mission, Gaia2022dr3} and \textit{Spitzer} data from the SAGE survey \cite{Bonanos2009, Bonanos2010}. This time, we only lost two targets (CPD-69 463 and LHA 120-S 83 without \textit{Spitzer} data), but were still left with 19 sources. 

\begin{figure}[H]
\begin{adjustwidth}{-\extralength}{0cm}
\centering
    \includegraphics[width=0.65\columnwidth]{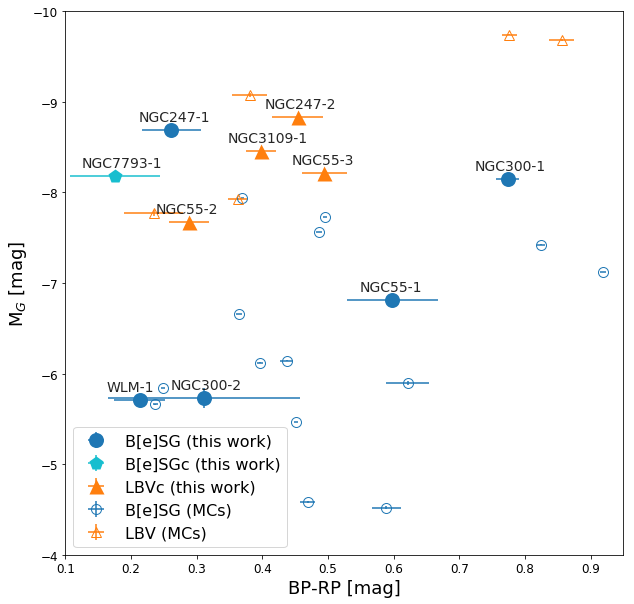}
    \includegraphics[width=0.65\columnwidth]{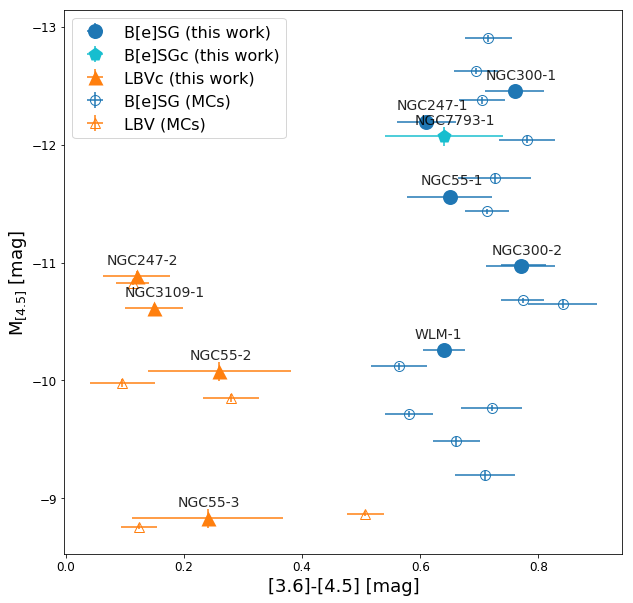}
  \end{adjustwidth}   
     \caption{(Left) The optical (\textit{Gaia}) CMD, plotting BP--RP vs. M$_{G}$ magnitude. We included all our sample and the MC sources from \cite{Kraus2019a} (except for two sources without a complete dataset in both \textit{Gaia} and \textit{Spitzer} surveys). (Right) The mid-IR (\textit{Spitzer}) CMD using the IR color [3.6]--[4.5] vs. M$_{[4.5]}$. In this case, there is a significant improvement in the separation between the two classes. The position of NGC7793-1 favors a B[e]SG nature (see text for more). 
     }   
     \label{f:cmds}   
\end{figure}

In Figure \ref{f:cmds} we present the optical (\textit{Gaia}) CMD, plotting BP--RP vs. M$_{G}$ band. We notice the lack of any correlation in the optical. 

In Figure \ref{f:cmds} we also present the mid-IR (\textit{Spitzer}) CMD, plotting [3.6]--[4.5] vs. M$_{[3.6]}$ band. The separation between the two classes becomes more evident in this case. The presence of hotter dusty environments becomes more significant for B[e]SGs, as they looked redder than LBVs (with a [3.6]--[4.5] range between 0.5 to 0.65 mag). They also tend to be much more luminous in the [3.6] than the LBVs. We  highlighted the position of NGC7793-1 in this plot. Although, from its spectrum alone, we could not determine a secure classification (due to issues with the obtained spectrum) it is located among the B[e]SGs of our sample and of the MCs. Therefore, we considered it a candidate B[e]SG. A future spectrum is needed to verify the existence of the [O~\textsc{i}]~$\lambda$6300 line, similar to the rest of the secure B[e]SGs in our sample. 

\begin{figure}[H]

  \includegraphics[width=0.7\columnwidth]{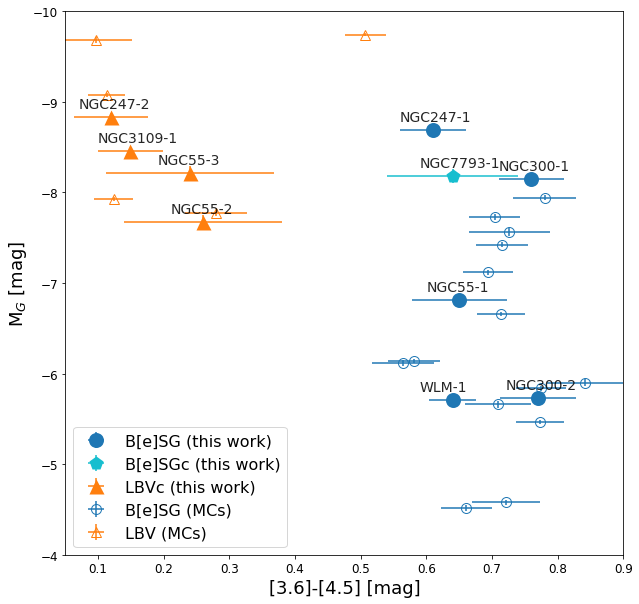}
    \caption{Similar to Figure \ref{f:cmds} but plotting the IR color [3.6]--[4.5] vs. the optical M$_{G}$ magnitude. Similar to the IR CMD we see relatively good separation between the two classes, with LBVs being brighter in the optical and less dusty compared to the B[e]SGs.  
    }
    \label{f:cmd_optir}
\end{figure}

We also tried to combine the optical and IR data in a CMD where we plot the [3.6]--[4.5] vs. M$_{G}$ magnitude (Figure \ref{f:cmd_optir}). The result was actually similar to the previous IR CMD (as the x-axis did not change). In this case, the plot can be more helpful, as the LBVs are populating the upper left part of the plot. Therefore, very bright optical sources with IR color up to $\sim$0.5~mag were most probably LBVs, while sources with color $>0.5$~mag would be B[e]SG (at almost any G magnitude).

\subsection{Metallicity Dependence of Populations}

\begin{figure}[H]

\includegraphics[width=\columnwidth]{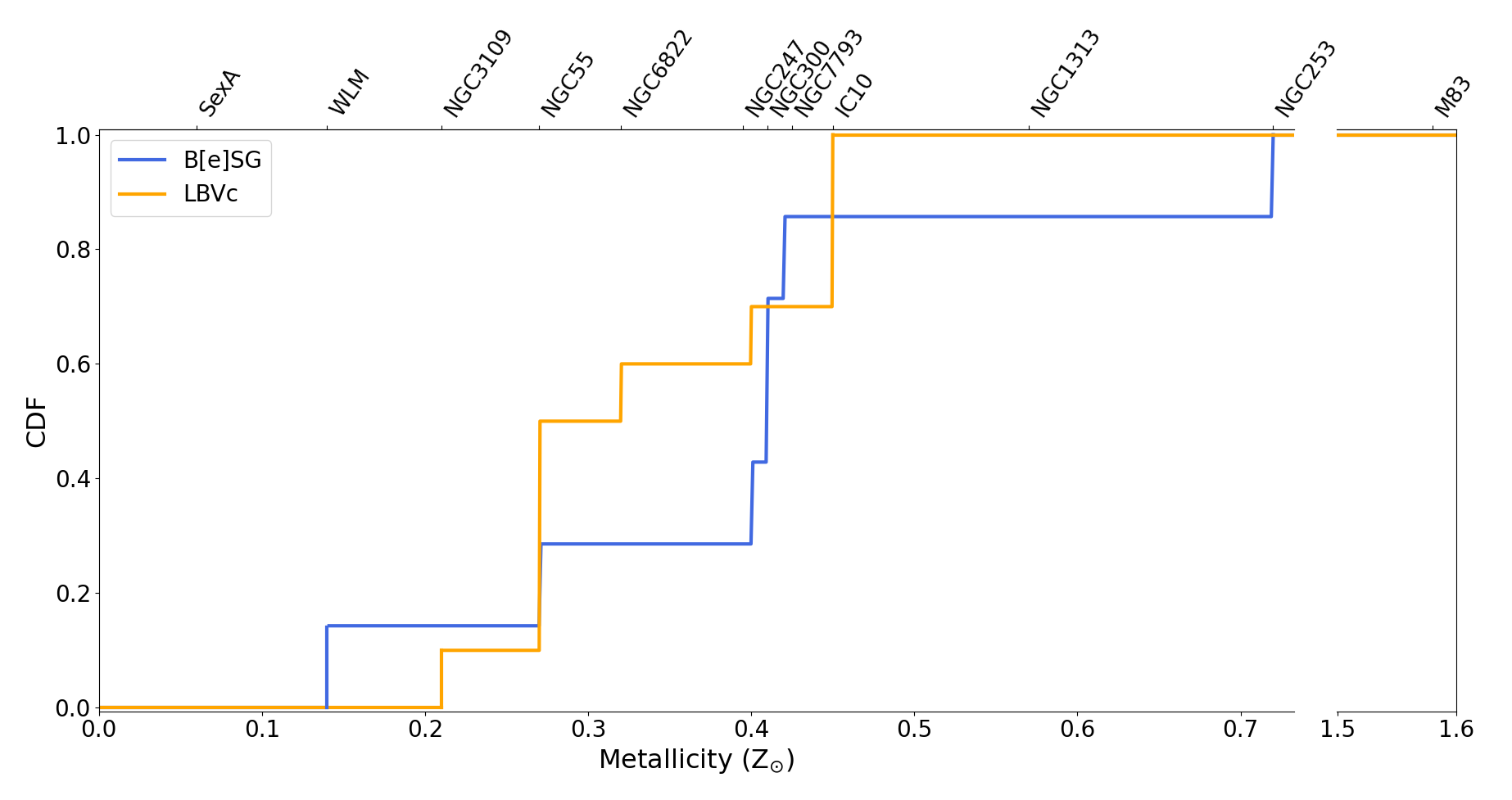}
    \caption{The cumulative distribution function of the B[e]SGs and LBVs (including candidates) from this work and the literature. We notice (for the first time) the presence of B[e]SGs in lower metallicity environments and the fact that the two populations are not totally different (see text for more). 
    }
    \label{f:distributions}
\end{figure}

In this section, we examine the populations of the two classes as a function of metallicity. For this, we plot the cumulative distribution function with metallicity (Figure~\ref{f:distributions}), considering all detected and known objects in our sample of galaxies. Namely, the numbers presented in Table~\ref{t:sources}, as well as the two LBVs in NGC55 \cite{Castro2008}, one in NGC 6822 and three in IC 10 \cite{Massey2007}, resulting in 7 B[e]SGs (including the NGC7793-1 candidate) and 10 LBVs in our sample of 12 galaxies.

We notice the presence of B[e]SGs at metallicity as low as $\sim$0.14 Z$_\odot$ (WLM). The current work is the first to detect these sources at such low metallicities. The population of LBVs begins at  $\sim$0.21 Z$_\odot$ (NGC 3109), and then increases steadily as we move towards higher metallicities. B[e]SGs presents an important step (increase) around $\sim$0.4 Z$_\odot$. In total, the two populations do not look significantly different. We have to be cautious interpreting this figure, however, due to the low number of statistics and completeness issues, as, for example, depending on the angle under which we observe a galaxy, we may not be able to fully observe its stellar content (e.g., NGC 253). 

\vspace{-9pt}

%%%%%%%%%%%%%%%%%%%%%%%%%%%%%%%%%%%%%%%%%%
\section{Conclusions}\label{sec5}

In this work, we report the detection of 6 secure B[e]SGs, 1 candidate B[e]SG, and 4 LBV candidates sources, of which 6 B[e]SGs and 2 LBVs are new discoveries. They are based on spectroscopic and photometric diagnostics, supplemented with RVs that are consistent with their host galaxies. By inspecting the available IR (2MASS, \textit{WISE}, \textit{Spitzer}) and optical (\textit{Gaia}) CMDs we find that the new sources are totally consistent with the loci of these populations from MCs. This adds further support regarding their natures. Building the cumulative distribution function of both populations with metallicity we notice the presence of B[e]SGs at environments with Z$\sim0.14$ Z$_{\odot}$, which increases the pool of extragalactic B[e]SGs and, especially, at lower metallicities. This is particularly important in order to investigate (with increased samples) these phases of massive stars. Since B[e]SGs and LBVs are among the classes with the most important episodic and outburst activities they provide valuable information on the role of episodic mass loss and insights into  stellar evolution in general. 

%%%%%%%%%%%%%%%%%%%%%%%%%%%%%%%%%%%%%%%%%%

\vspace{6pt} 

%%%%%%%%%%%%%%%%%%%%%%%%%%%%%%%%%%%%%%%%%%

\authorcontributions{Conceptualization, G.M. and A.Z.B.; Funding acquisition, A.Z.B.; Investigation, G.M., S.d.W., A.Z.B., G.M.-S. and E.C.; Methodology, G.M., S.d.W. and F.T.; Software, F.T. and G.M.-S.; Supervision, A.Z.B.; Visualization, G.M. and S.d.W.; Writing---original draft, G.M., S.d.W. and A.Z.B.; Writing---review \& editing, G.M., S.d.W., A.Z.B., F.T., G.M.-S. and E.C. All authors have read and agreed to the published version of the manuscript.}

\funding{This research was funded by the European Research Council (ERC) under the European Union’s Horizon 2020 research and innovation programme (Grant agreement No. 772086).} 

\dataavailability{Photometry and 1D extracted spectra will become available through the VizieR/CDS catalog tool.} 

\acknowledgments{GM acknowledges feedback from Francisco Najarro and Michaela Kraus. 
Based on observations collected at the European Southern Observatory under the ESO programme 105.20HJ and 109.22W2. Based on observations made with the Gran Telescopio Canarias (GTC), installed at the Spanish Observatorio del Roque de los Muchachos of the Instituto de Astrofísica de Canarias, on the island of La Palma (programme GTC83/20A). This work was (partly) based on data obtained with the instrument OSIRIS, built by a Consortium led by the Instituto de Astrofísica de Canarias in collaboration with the Instituto de Astronomía of the Universidad Autónoma de México. OSIRIS was funded by GRANTECAN and the National Plan of Astronomy and Astrophysics of the Spanish Government. This work was based, in part, on observations made with the \textit{Spitzer} Space Telescope, which is operated by the Jet Propulsion Laboratory, California Institute of Technology, under a contract with NASA. This work made use of data from the European Space Agency (ESA) mission \textit{Gaia} (\url{https://www.cosmos.esa.int/gaia}), processed by the \textit{Gaia} Data Processing and Analysis Consortium (DPAC, \url{https://www.cosmos.esa.int/web/gaia/dpac/consortium}). Funding for the DPAC was provided by national institutions, in particular, the institutions participating in the \textit{Gaia} Multilateral Agreement. Based on observations made with ESO Telescopes at the La Silla or Paranal Observatories under programme ID(s) 179.A-2010(A), 179.A-2010(B), 179.A-2010(C), 179.A-2010(D), 179.A-2010(E), 179.A-2010(F), 179.A-2010(G), 179.A-2010(H), 179.A-2010(I), 179.A-2010(J), 179.A-2010(K), 179.A-2010(L), 179.A-2010(M), 179.A-2010(N), 179.A-2010(O) (regarding VISTA Hemisphere Survey). This publication made use of data products from the Two Micron All Sky Survey, which is a joint project of the University of Massachusetts and the Infrared Processing and Analysis Center/California Institute of Technology, funded by the National Aeronautics and Space Administration and the National Science Foundation. This publication  used data products from the Wide-field Infrared Survey Explorer, which is a joint project of the University of California, Los Angeles, and the Jet Propulsion Laboratory/California Institute of Technology, funded by the National Aeronautics and Space Administration.
This work used Astropy \url{http://www.astropy.org}: a community-developed core Python package and an ecosystem of tools and resources for astronomy \citep{astropy:2013, astropy:2018, astropy:2022}, NumPy (\url{https://numpy.org/}; \cite{numpy2020}), and matplotlib (\url{https://matplotlib.org/}; \cite{matplotlib})
}

\conflictsofinterest{The authors declare no conflict of interest.} 

%%%%%%%%%%%%%%%%%%%%%%%%%%%%%%%%%%%%%%%%%%
\abbreviations{Abbreviations}{
The following abbreviations are used in this manuscript:\\

\noindent 
\begin{tabular}{@{}ll}
B[e]SG & B[e] Supergiant\\
CCD & Color-Color Diagram \\
CMD & Color-Magnitude Diagram \\
CSE & circumstellar environment \\
LBV & Luminous Blue Variable\\
MC & Magellanic Cloud\\
SNR & Signal to Noise Ratio\\
RSG & Red Supergiant\\
RV & Radial Velocity
\end{tabular}
}

%%%%%%%%%%%%%%%%%%%%%%%%%%%%%%%%%%%%%%%%%%
\begin{adjustwidth}{-\extralength}{0cm}

\reftitle{References}
%\bibliography{your_external_BibTeX_file}
\bibliography{refs.bib}

%%%%%%%%%%%%%%%%%%%%%%%%%%%%%%%%%%%%%%%%%%
\end{adjustwidth}
\end{document}